\documentclass[a4paper]{article}
\usepackage{epsfig,amsmath,amsfonts,amssymb,setspace,multirow,textcomp}
\usepackage[T1]{fontenc}
\usepackage{hyperref}
\makeatletter
\renewcommand{\@oddhead}{\textit{Advances in Astronomy and Space Physics} \hfil}
\renewcommand{\@evenfoot}{\hfil \thepage \hfil}
\renewcommand{\@oddfoot}{\hfil \thepage \hfil}
\makeatother

\renewenvironment{thebibliography}[1]{\begin{oldthebibliography}{#1}\setlength{\parskip}{0ex}\setlength{\itemsep}{0ex}}{\end{oldthebibliography}}

\usepackage{float} 
\newcommand{\mysp}{}\def\mysp/{}

\begin{document}
\fontsize{11}{11}\selectfont 
\title{Ammonia in circumstellar environment of V Cyg}
\author{\textsl{B.~Etma{\'n}ski$^{1}$, M.\,R.~Schmidt$^{1}$, R.~Szczerba$^{1}$}}
\date{\vspace*{-6ex}}
\maketitle
\begin{center} {\small $^{1}$Nicolaus Copernicus Astronomical Center, Polish Academy of
Sciences, ul. Rabia{\'n}ska 8, 87-100 Toru{\'n}, Poland\\}
\end{center}

\begin{abstract}
The HIFI instrument on board of the Herschel Space Observatory (HSO) has been very successful in detecting
 molecular lines from circumstellar envelopes around evolved stars, like massive red supergiants,
 Asymptotic Giant Branch (AGB) and post-AGB stars, as well as planetary nebulae. Among others, ammonia has
been found in circumstellar envelopes of C-rich AGB stars in amounts that significantly exceeded
theoretical predictions for C-rich stars. Few scenarios have been proposed to resolve this problem: formation
of ammonia behind the shock front, photochemical processes in the inner part of the envelope partly
transparent to UV background radiation due to the clumpy structure of the gas, and formation of ammonia on
dust grains. Careful analysis of observations may help to put constraints on one or another mechanism of
ammonia formation. 
Here, we present results of the non-LTE radiative transfer modeling of ammonia
transitions including a crucial process of radiative pumping via v$_2$ = 1 vibrational band (at $\sim$10\,$\mu$m)
for V Cyg.
Only ground-based
ammonia transition NH$_{3}$ J = 1$_{0}$ - 0$_{0}$ at 572.5 GHz has been observed by HIFI. Therefore, to determine abundance of ammonia 
we estimate a photodissociation radius of NH$_{3}$ using chemical model of the envelope consistent with dust grain properties concluded from the spectral energy distribution. \\[1ex]
\end{abstract}

\section*{\sc introduction}
\indent \indent  Ammonia ($NH_{3}$) was discovered in space in 1968 \cite{cheung1968} and its first transition was detected at $23$ $GHz$ with radiotelescope. Ammonia is mostly observed in interstellar medium \cite{ho1983}, and commonly in star forming regions (see e.g. \cite{harju1993}, \cite{jijina1999},
\cite{wienen2012}). However, ammonia is also detected in the cricumstellar envelopes of evolved stars (see, e.g. \cite{schmidt2016} for analysis of ammonia in Asymptotic Giant Branch (AGB) star CW\,Leo).
The Asymptotic Giant Branch stage of stellar evolution
is characteristic for stars with initial mass between 1-8 M$_{\odot}$, that is for stars with low-to-intermediate initial mass, and is characterized by intense mass-loss processes. The AGB stars can be divided into two main types: 
C-rich and O-rich ones. The C-rich AGB  stars contain more carbon than oxygen atoms, i.e. C/O > 1. V Cyg itself belongs to this group of objects. 
V Cyg is Mira type variable with period of about 417 days, C6 spectral type 
\cite{wallerstein1998} and a distance of 420 pc \cite{whitelock2008}. The central star
has low effective temperature (2000-3000K). The star in this phase of evolution has very weak 
gravitational force in the stellar atmosphere and changes periodically its radius, temperature and luminosity. 
Pulsations are responsible for the dust formation in the outer layers of the atmosphere, and then radiation 
pressure on dust can initiate the ejection of material form the atmosphere into space. 
Mater ejected from the star forms Circumstellar Envelope (CSE) around the star, where 
new molecules, like for example ammonia, are formed. The observed abundance of the NH$_{3}$ is
several magnitudes larger than models of stellar atmosphere for AGB stars predict (e.g. \cite{johnson1982}). The 
formation process of ammonia is unknown. The two scenarios have been proposed for ammonia formation in CSE.
In the first scenario ammonia may be formed  by passage of shocks. In this scenario 
N$_{2}$ molecules have been dissociated and atoms of nitrogen are available to form the ammonia 
molecule \cite{willacy1998}. Unfortunately, the last calculations have been shown that this process change
the abundance of NH$_{3}$ only marginally \cite{gobrecht2016}. The second mechanism is dissociation of N$_{2}$
by the UV radiation because of the CSE is inhomogeneous and the UV radiation may penetrate almost
through the whole envelope \cite{decin2010}.

In Section 2 we present the description of the molecular model of ammonia. The observations and data
reductions are described in Section 3. Section 4 is 
devoted to description of modelling process of ammonia lines
in CSE of V\,Cyg. At the last, in Section 5 we discuss 
results of our modelling.

\section*{\sc Model of ammonia}

\indent \indent  Ammonia molecule exists in two species: ortho and para. These two species
differ in the orientations of the three hydrogen spins.
Symmetry properties of the ro-vibrational (including inversion) levels are
described by the symmetry group $D_{3h}$ (C$_{3v}$ if the inversion mode is
neglected).
Because of the Pauli exclusion principle, the ro-vibrational levels must belong to
$A'$ and $A"$ (para states) and 
$E'$ and $E"$ (ortho states) representations of $D_{3h}$ \cite{bunker2006}. 
Ammonia oscillates in six vibrational modes, of which two pairs are
degenerate, and which are described by the four quantum vibrational numbers
($\nu_{1}$, $\nu_{2}$, $\nu_{3}^{l_{3}}$, $\nu_{4}^{l_{4}}$)
and the two vibrational angular momentum l$_3$, and l$_4$.

Ammonia molecule has inversional nature. It is because the potential
energy surface has double minimum corresponding to two positions of the nitrogen atom
on opposite sides of the plane defined by hydrogen atoms. The molecule
may change rapidly its geometry, because its potential barrier between the two minima is small and 
equal $2009$ $cm^{-1}$ ($6.4 \times 10^{-14}$ $erg$). In this case energy levels are not
strongly separated and wavefunctions mix together resulting in two kinds of levels: symmetric (s, +) and 
antisymmetric (a, -) \cite{bernath2005}.

For calculations of excitation of a molecule in circumstellar envelope one
needs the input file containing the list of adopted energy levels, list of
radiative transitions between energy levels and the set of collisional rates.
For limited number of levels the input file is available in the database
LAMDA (\url{https://home.strw.leidenuniv.nl/~moldata}, \cite{denby1988}, \cite{maret2009}). 
Hereby, we followed Schmidt et al. (2016) and the set of levels
extracted from the theoretical EXOMOL database (\url{http://exomol.com},\cite{tennyson2016}) 
has been used.
The advantage of using computed database instead of laboratory is its completeness.

\section*{\sc Data}

\indent \indent Observations of V Cyg were obtained with the HIFI (\textit{Heterodyne Instrument for the Far Infrared}) instrument on board of the HSO (\textit{Herschel Space Observatory}) in the point mode.
The observations were part of the HIFISTARS Guaranteed Time Key Program (Proposal Id: 
KPGT\_vbujarra\_1, PI: V.Bujarrabal). Details of available observations are presented in 
Table \ref{tab1}.  For each observations we give spectroscopic designation, 
frequency of transition in GHz, band of the HIFI instrument, energy of the upper level in K, observational
identification of, date of observation, the phase of observation, half power beam with (HPBW) of 
HIFI instrument, and the main beam efficiency ($\eta_{mb}$) ({Mueller et al.
2014\footnote{\url{http://hertwiki.esac.esa.int/twiki/pub/Public/HifiCalibrationWeb/HifiBeamReleaseNote_Sep2014.pdf}}).
The observational phase of variability was determined relative to the reference data of maximum,
selected from the visual variability curve (see below) ($\phi =$ 0 for Julian Date 2450330), and assuming
variability period of 417 days \cite{samus2009}).
Only the ground rotational transition of ortho-NH$_{3}$ was recorded in two instrumental settings
observed at the same time.
The reduction of data was conducted using HIPE (v.15.0.1)
software\footnote{\url{https://www.cosmos.esa.int/web/herschel/hipe-download}}. 
The reduction process was limited
to extraction of data from level 2, tracing continuum and removing baseline, averaging of two H and V
polarizations, and final averaging of two settings. For comparison with theoretical models, the antenna
temperature was converted to the main beam temperature using main beam coefficients. The final
profile of the 1$_{0}$-0$_{0}$ ortho-ammonia line is shown in Figure~\ref{fig1}
on a Doppler velocity scale relative to the Local Standard of Rest V$_{LSR}$.
The root mean square (r.m.s.) in the continuum amounts to 3~mK. 
The uncertainty of the flux is expected to be less than 10 percent \cite{roelfsema2012}.

For the analysis of variability we used data available in AAVSO database\footnote{https://www.aavso.org/}.
AAVSO database provides a very long sequence of photometric observations in visual covering
both HIFI and ISO periods of observations. Sequence of data covering ISO observations is shown in
Figure~\ref{fig2}.

The main source of data for the Spectral Energy Distribution (SED) modelling is atlas of 
ISO (Infrared Space Observatory)
observations available on \url{https://users.physics.unc.edu/~gcsloan/library/swsatlas/aot1.html}
(\cite{kraemer2002}, \cite{sloan2003}). This catalogue includes 1239 full-scan AOT1 spectra from
Short Wavelength Spectrometer (SWS). These spectra contain photometric scan between about 2.4 to
45.4 $\mu$m for each object. All spectra were reduced and ready to use for modelling \cite{sloan2003}.
In collection of spectra we found six spectral scans of our object (Table \ref{tab2}).
One spectral scan was observed at variability phase close to that of HIFI observations (ID = 69500110)
and was used for modelling of the SED (Figure \ref{fig3}). 
Photometry was updated with photometrical measurements  of V\footnote{\url{http://simbad.u-strasbg.fr/simbad/sim-ref?querymethod=bib&simbo=on&submit=submit+bibcode&bibcode=2002yCat.2237....0D}},
J, H and K bands (\cite{cutri2003}, date of observation: 2000-05-04, JD date: 2451668.9397, 
    phase: $\phi=$ 0.2)\footnote{\url{http://vizier.u-strasbg.fr/viz-bin/VizieR?-source=II/246}} and 
with IRAS measurements for 12, 25, 60 and 100 $\mu$m\footnote{\url{http://vizier.u-strasbg.fr/viz-bin/VizieR-5?-out.add=.&-source=II/125/main,II/126/sources&IRAS==20396\%2b4757\%2a}}
(in this case the flux reported in the catalogue
is average value of the IRAS observations at each band, and the phase is unknown
\footnote{https://lambda.gsfc.nasa.gov/product/iras/colorcorr.cfm}).

\begin{table}
 \centering
 \caption{The characteristics of HIFI/\textit{Herschel} observations.}\label{tab1}
 \vspace*{1ex}
 \begin{tabular}{ccccccccc}
  \hline
  Transition & Frequency & Band & E$_{u}$ & \textit{Herschel} & Obs. date & Phase & HPBW & $\eta_{mb}$ \\
            & (GHz)      &      & (K)     & OBSID             &           &       &      &             \\
  \hline
  1$_{0}$(s) - 0$_{0}$(a) & 572.498 & 1b & 29 & 1342199153 & 2010-23-06  & 0.1  & 37.5" & 0.62 \\
                          &         &    &    & 1342199154 & 2010-23-06  & 0.1  & 37.5" & 0.62 \\
 \hline 
 \end{tabular}
\end{table}

\begin{table}
 \centering
 \caption{The parameters of the ISO SWS for V Cyg.}\label{tab2}
 \vspace*{1ex}
 \begin{tabular}{ccccc}
  \hline
  Source name & ID & Date & Time of observation (s) & Phase \\
  \hline
  V Cyg & 08001855 & 1996-05-02 & 1044 & 0.49\\
        & 42100111 & 1997-10-01 & 1912 & 0.31\\
        & 42300307 & 1997-12-01 & 1913 & 0.31\\
        & 51401308 & 1997-13-04 & 1912 & 0.53\\
        & 59501909 & 1997-03-07 & 1912 & 0.73\\
        & 69500110 & 1997-10-10 & 1912 & 0.99\\ 
 \hline 
 \end{tabular}
\end{table}

\section*{\sc Modelling}

\indent \indent The first step in modelling was preparation of the model of the dusty envelope of V~Cyg.
For that purpose, we used MRT code for a dust radiative transfer \cite{szczerba1997}.
In SED modelling we assumed that the amplitude of luminosity variation is
about 40 \% 
relative to the average value deduced
from the period-luminosity relation \cite{whitelock2008} 
of 6600 L$_{\odot}$.
Such an amplitude was deduced for the similar case for the best examined C-rich
star - CW Leo \cite{menshchikov2001}. Hence, in the phase of maximum the obtained luminosity 
equals 9200 L$_{\odot}$. 
The optimal model of the dusty envelope reproduces well observed SED for V Cyg both in the maximum phase
(presented on Figure~\ref{fig1}) and in the minimum phase. The adopted parameters of the central source
and of the dust are presented in Table~\ref{tab3}.

\begin{table}[H]
 \centering
 \caption{The parameters for model of V Cyg.}\label{tab3}
 \vspace*{1ex}
 \begin{tabular}{cccc}
  \hline
  \hline
  RA (2000)     & 20 41 18.27                                      & Luminosity (mean) & 6600$^2$ L$_{\odot}$\\
  Dec (2000)    & +48 08 28.94                                     & Luminosity (max)  & 9200$^4$ L$_{\odot}$\\
  Period        & 417 days$^1$                                     & Luminosity (min)  & 4000$^4$ L$_{\odot}$\\
  Distance      & 420 pc$^2$                                       & T$_{eff, max}$    & 2500$^4$ K\\
  V$_{exp}$     & 11.5 km s$^{-1}$ $^3$                            & T$_{eff, min}$    & 2000$^4$ K\\
  V$_{LSR}$     & 13.5 km s$^{-1}$ $^6$                            & f(ortho-NH$_{3}$) & 3.3 $\times$ 10$^{-6}$ $^4$\\
  R$_{phot}$    & 1.5 $\times$ 10$^{16}$ cm                        & f(NH$_{3}$)       & 6.6 $\times$ 10$^{-6}$ $^4$\\
  M$_{loss}$    & 1.7 $\times$ 10$^{-6}$ M$_\odot$ yr$^{-1}$ $^5$  &                   & \\
 \hline
 \hline
 \end{tabular}
 \\$^1$ \cite{samus2009}, $^2$ \cite{whitelock2008}, $^3$ Derivatives from CO lines, $^4$ SED modelling,
 $^5$\cite{shoier2013}, $^6$ \cite{massalkhi2018}
\end{table}

The optical properties of the dust 
obtained
from the 
SED
modelling were used to calculate the
photodissociation radius of ammonia. For that purpose, we used CSENV (\textit{CircumStellar ENVelopes})
code for the calculations of chemistry in expanding envelope \cite{glassgold1986}. The chemical contents
of the input was restricted to ammonia and products of its photodissociation. 
The photodissociation radius, R$_{ph}$ is defined as the radius at which the initial
abundance of ammonia drops twice. In our calculations we have used the latest cross-section rates of ammonia
molecules \cite{heays2017}.
In case of V~Cyg we obtained value of 1.5 $\times$ 10$^{16}$ cm (for the mass loss
rate $\dot{M} =$ 1.7 $\times$ 10$^{-6}$ M$_{\odot}$ yr$^{-1}$ and expansion velocity V$_{exp} =$ 11.5 km
s$^{-1}$). Applying the same approach to analysed earlier CW Leo \cite{schmidt2016}, we found
theoretical R$_{ph}$ equal to 4 $\times$ 10$^{16}$ cm, 
close to the value 
obtained observationally
of 3.5 $\times$ 10$^{16}$ cm for the rotational transitions of ortho-NH$_{3}$ (2.0 $\times$ 10$^{16}$ cm
suggested by inversional transitions). We estimate that the uncertainty of the theoretical value
may be of factor 2.

For modelling of ammonia lines we used a radiative transfer code MOLEXCSE (\textit{Molecular Line EXcitation in CircumStelar Envelope}) code. This code solves the non-LTE radiative transfer in molecular lines
in presence of the continuum. The code reproduces well flux in continuum for the absorption, scattering and emissivity provided by MRT code. 
The density structure of the gas envelope is determined by the assumed mass loss rate and constant expansion velocity.
The gas kinetic temperature distribution in the envelope of V~Cyg was
determined by finding the best fit to CO emission lines including both radio, 
milimeter and sub-milimeter HIFI transitions (\cite{neufeld2010}). 

The model of ortho-ammonia molecule includes 172 levels (all the ground vibrational 
and the first excited bending mode $\nu_2$=1 levels with J $=<$ 15), 
1579 radiative transitions and collisional excitation by H$_{2}$. For the details of
treatment of collisional rates see \cite{schmidt2016}.
Inclusion of the rovibrational transitions (about 10 $\mu$m) is of key importance 
for the proper modelling of excitation of ammonia in circumstellar envelope 
(see e.g. \cite{schmidt2016}).

\begin{figure}[!h]
\centering
\begin{minipage}[t]{.45\linewidth}
\centering
\epsfig{file = 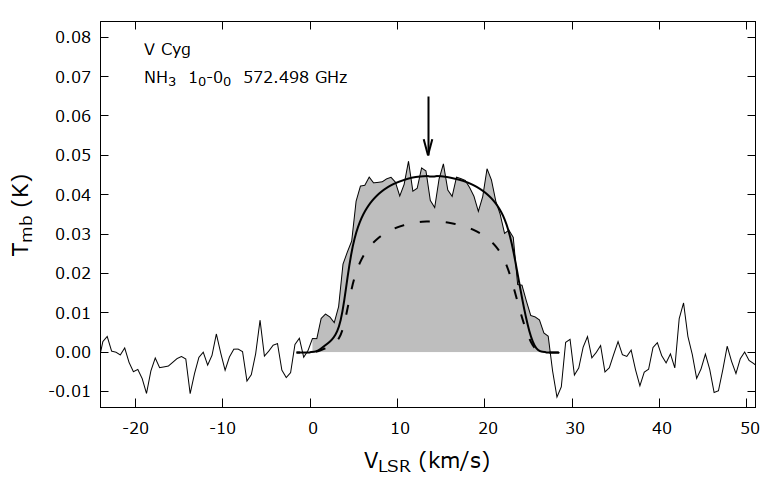,width = .99\linewidth}
\caption{
The averaged line profile of NH$_{3}$ obtained from HIFI/\textit{Herschel}
observations.
The black solid line shows the brightness for the model of ammonia in maximum phase of variability of VCyg,
the black dashed line - in minimum phase of variability. The vertical arrow indicates 
the LSR velocity of the source\cite{massalkhi2018}.}\label{fig1} 
\end{minipage}
\hfill
\begin{minipage}[t]{.45\linewidth}
\centering
\epsfig{file = 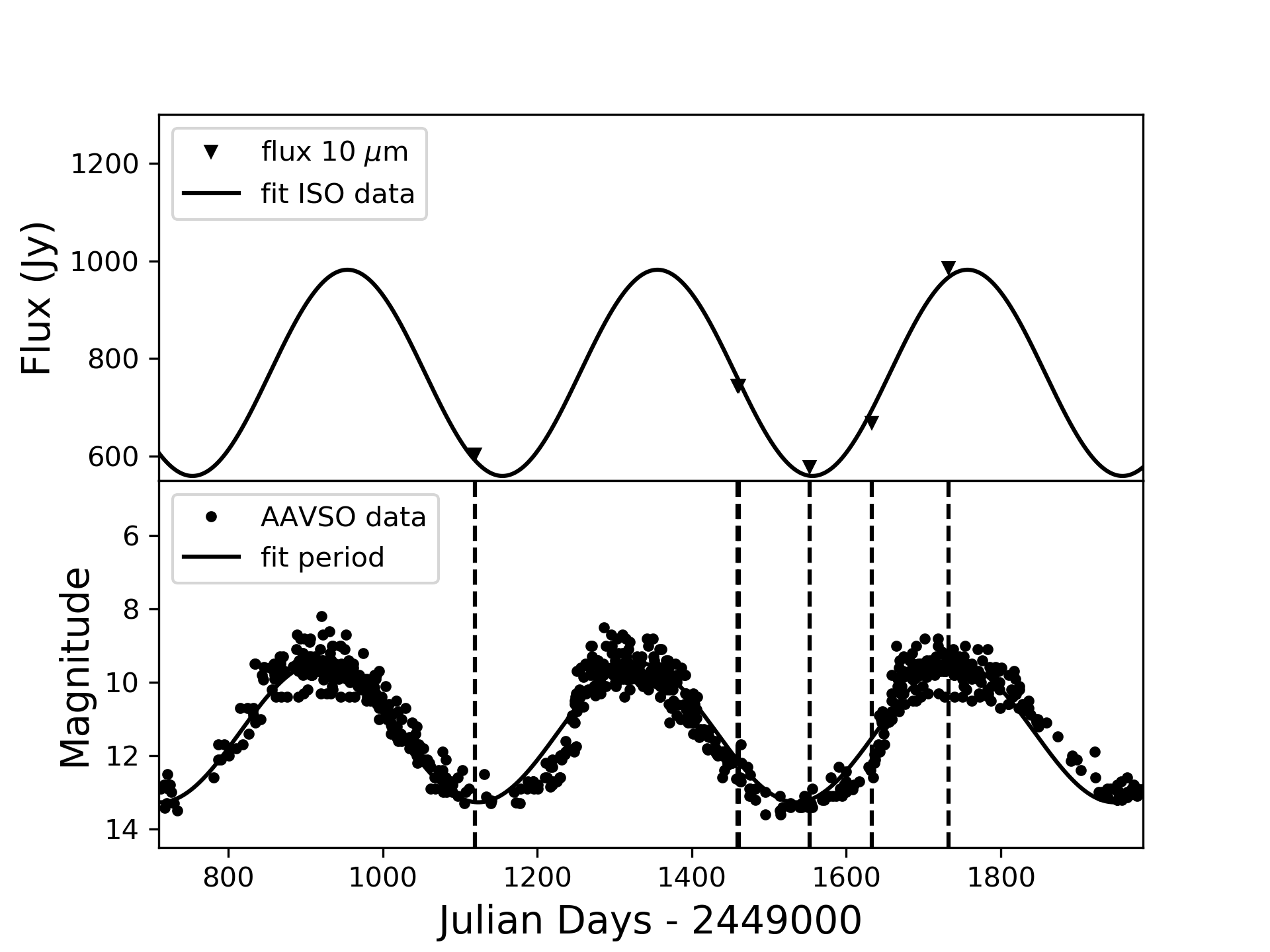,width = .99\linewidth}
\caption{Variability of V~Cyg. The upper diagram shows the measurements flux at 10 $\mu$m from
ISO observations (triangle points) and fit sine function to show how flux 
changed with the period of variability. The bottom diagram presents AAVSO data (circle points,
visual observations, AAVSO (American Association of Variable Stars Observers)),
and determinations of period
(solid line). The broken vertical line
presents positions of the ISO observations for V Cyg. The shift between
amplitude of this two sets of data is equal about 0.07.}\label{fig2}
\end{minipage}
\end{figure}

\begin{figure}[!h]
\centering
\begin{minipage}[t]{.45\linewidth}
\centering
\epsfig{file = 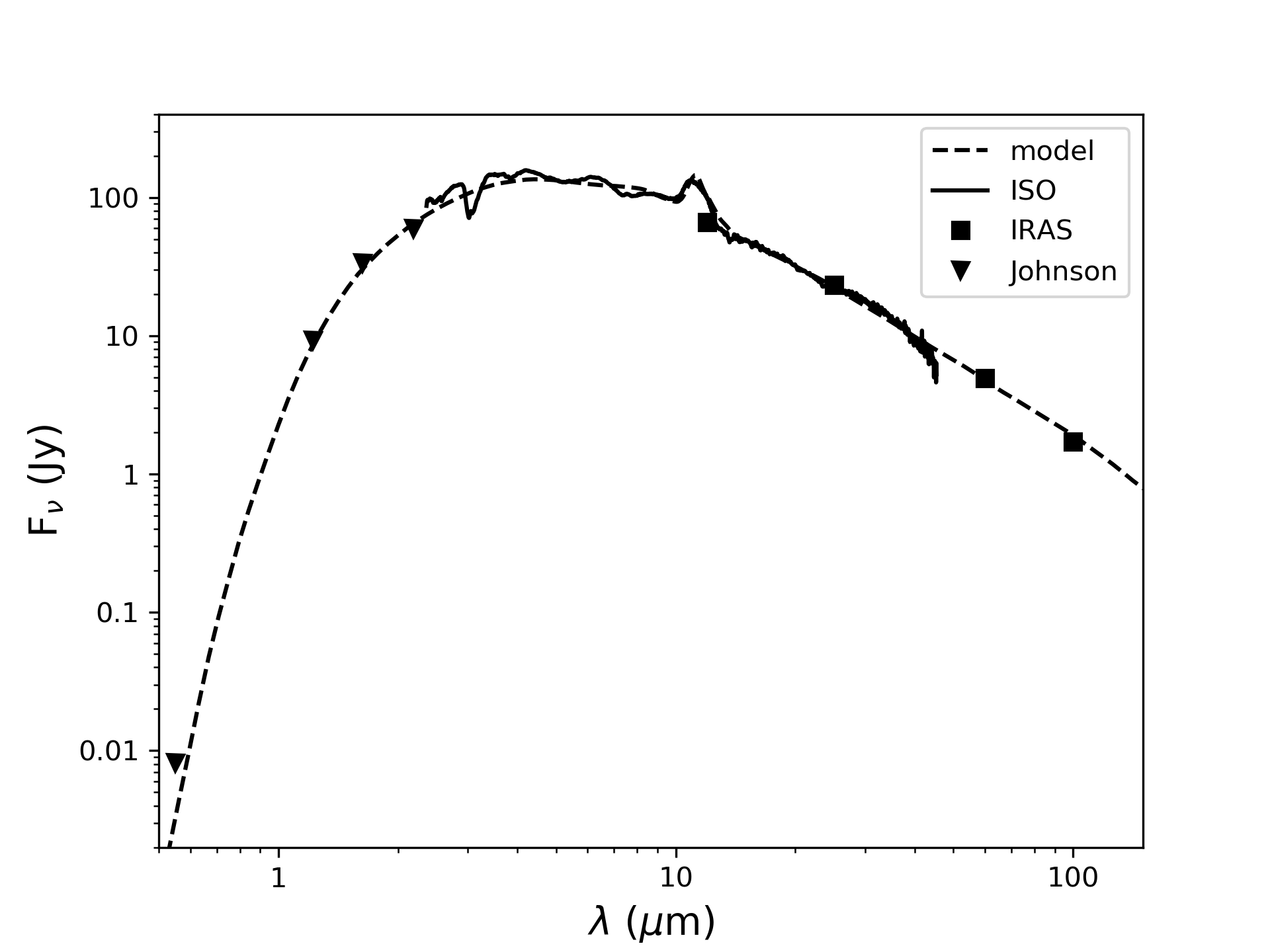,width = .99\linewidth}
\caption{
The SED (\textit{Spectral Energy Distribution}).
Triangle points shows measurements flux for V, J, H and K bands,
square points 
- IRAS measurements for 12, 25, 60 and 100 $\mu$m, the solid line - ISO
measurements between 2.36 - 45.365 $\mu$m (ID: 69500110, phase: $\phi =$ 0.99).
The 
dashed line shows model of SED.}\label{fig3} 
\end{minipage}
\hfill
\begin{minipage}[t]{.45\linewidth}
\centering
\epsfig{file = 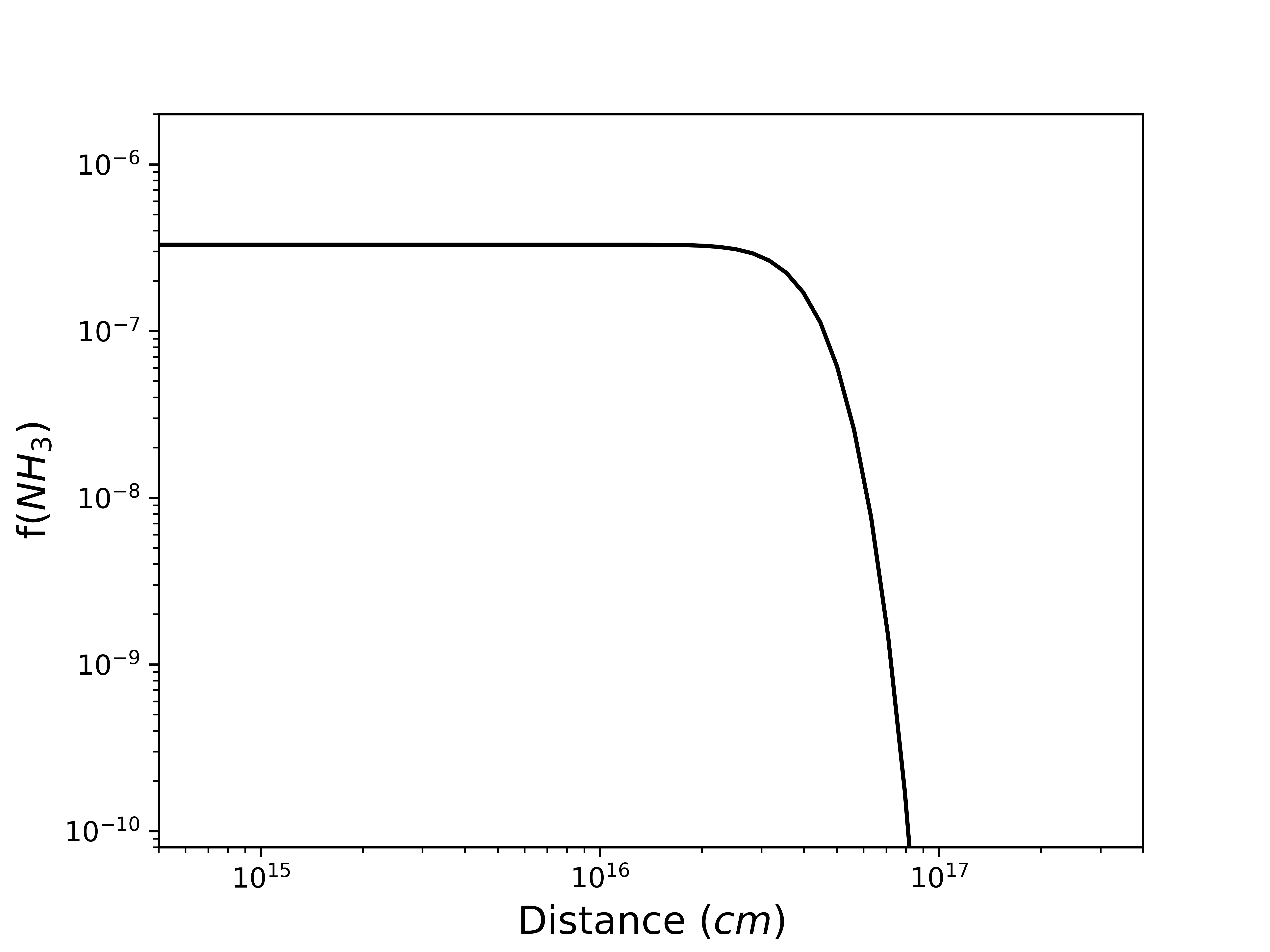,width = .99\linewidth}
\caption{
The distribution of NH$_{3}$ molecule in CSE around V Cyg. 
This model was been calculated by CSENV code \cite{glassgold1986}. This distribution
have been calculated for the model of dust obtain for our SED model and the latest 
cross-section of ammonia molecule \cite{heays2017}.}\label{fig4}
\end{minipage}
\end{figure}

\section*{\sc Result and conclusion}



For adopted parameters the abundance of ortho-NH$_{3}$ molecules in V~Cyg star in
maximum phase which reproduces at best observed line profile amounts to
3.3 $\times$ 10$^{-7}$ relative to the molecular hydrogen. 
The resulting theoretical profile is shown on Figure~\ref{fig3} with the black line.
The computed profile seems to reproduce very well almost flat top shape of the observed transition.

To illustrate expected variability in line profile, we computed the theoretical profile
of 1$_{0}$-0$_{0}$ transition also 
from the minimum phase, using abundance determination
in the maximum phase. The resulting profile is presented on Figure~\ref{fig3} by means of the dotted line.
In that case we used ISO data in the minimum phase (ID: 08001855, phase: $\phi =$ 0.49). 
For the model of the SED in the phase of minimum, we adopted central star luminosity to be equal 4000 
L$_{\odot}$ and the temperature to be 2000 K. 
The brightness of the 572.498 GHz line is then reduced by 40~\% relative to its 
maximum.

No observations of para-ammonia are available for V~Cyg. If ortho-to-para ratio corresponds to
the equilibrium conditions at higher temperatures, i.e. it is equal to one, then the total amount of ammonia
in the envelope of V~Cyg is 6.6 $\times$ 10$^{-7}$. The abundance of ammonia in the archetypical
C-rich AGB star IRC+10216 amounts 6 $\times$ 10$^{-8}$ \cite{schmidt2016} for the adopted there mass loss rate of 3.25 $\times$ 10$^{-5}$ M$_{\odot}$ yr$^{-1}$. Thus, within uncertainties one may conclude that
mass production rate of ammonia in envelopes of both objects is similar despite significantly different
gas densities. If photochemical reactions would be responsible for the production of ammonia 
in the envelopes of C-rich AGB stars, then higher abundance of ammonia in V~Cyg could be explained
by the higher probability that UV photons may penetrate into inner parts of the envelope while its density is smaller. The analysis of other 6 objects in the sample, now under consideration, will reveal if that expectations will be confirmed.


\section*{\sc acknowledgement}
\indent \indent 
We acknowledge support from the grant 2016/21/B/ST09/01626
of the National Science Centre, Poland. 


\end{document}